# Correction of beam wander for a free-space quantum key distribution system operating in urban environment

**Alberto Carrasco-Casado, Natalia Denisenko, Veronica Fernandez**
Spanish National Research Council (CSIC)
Institute of Physical and Information Technologies (ITEFI)
Serrano 144, 28006 Madrid, Spain

**Abstract**. Free-space quantum key distribution links in urban environment have demanding operating needs, such as functioning in daylight and under atmospheric turbulence, which can dramatically impact their performance. Both effects are usually mitigated with a careful design of the field of view of the receiver. However, a trade-off is often required, since a narrow field of view improves background noise rejection but it is linked to an increase in turbulence-related losses. In this paper, we present a high-speed automatic tracking system to overcome these limitations. Both a reduction in the field-of-view to decrease the background noise and the mitigation of the losses caused by atmospheric turbulence are addressed. Two different designs are presented and discussed, along with technical considerations for the experimental implementation. Finally, preliminary experimental results of beam wander correction are used to estimate the potential improvement of both the quantum bit error rate and secret key rate of a free space quantum key distribution system.

**Address all correspondence to:** Alberto Carrasco-Casado, Spanish National Research Council (CSIC), Institute of Physical and Information Technologies (ITEFI), Serrano 144, Madrid, Spain, 28006; Tel: +34 915618806 ext. 446; E-mail: alberto.carrasco@iec.csic.es

## 1 Introduction

Quantum key distribution (QKD) [1], and quantum cryptography in general, has become a new paradigm in data protection. The laws of quantum mechanics offer a theoretically-secure alternative for data communications over conventional methods, since the presence of an eavesdropper can be uniquely detected in the process of key sharing over an insecure channel.

Free-space QKD has been extensively aimed to satellite communications with the main efforts concentrating in achieving long distances to proof its feasibility [2]. However, short distance (inter-city range) free-space QKD links in urban areas may also offer some advantages over optical fiber, such as flexibility of installation and portability. Unlike optical fiber-based systems, free-space-based links can be easily transported to different locations if required. In this context, free-space QKD could be of interest to organizations such as financial, governmental and military institutions within the same city. These links may also be integrated to fiber-optic metropolitan networks and provide higher bandwidth when affected by poor connectivity.

Nevertheless, for QKD to be a realistic alternative, it has to operate at high speed, in daylight conditions and under atmospheric turbulence, which tends to be stronger in urban environments. A suitable tracking subsystem capable of fine correction of turbulent effects is therefore required. However, even though tracking techniques are common in traditional free-space optical communications, it is not so commonly used in QKD, where tracking is usually limited to maintaining a coarse alignment of the link. As a consequence turbulence ends up adding considerable losses to the optical link [3] resulting in a significant decrease of the key rate.

In this paper, we will analyze two automatic high-speed tracking techniques implemented to compensate beam wander caused by atmospheric turbulence for a QKD system described in [4].

## 2  The free-space QKD system

The QKD system implements the B92 polarization-encoding protocol [5]. A schematic diagram of the system can be seen in fig. 1. The emitter uses a GHz pulse pattern generator to provide a pre-programmed electronic sequence that feeds a high speed driver, which in turn controls two Gbps vertical-cavity surface-emitting lasers (VCSEL), represented by $V_0$ and $V_1$. The $\lambda \sim 850$ nm emission of each VCSEL, linearly polarized and set at a relative angle of 45˚, is used to encrypt the binary data of the cryptographic key. This is usually referred to as the *quantum signal*, which due to the current immaturity of quantum repeaters cannot be amplified. Moreover this signal is heavily attenuated to a single photon regime for security purposes and therefore losses in the transmission channel and the receiver should be minimized. This is also the reason why timing synchronization of emitter and receiver is performed by a third VCSEL ($V_{SYNC}$) emitting at $\lambda \sim 1550$ nm instead of using a fraction of the quantum signal; thus avoiding extra losses of the quantum signal. This emission is not attenuated to a single photon regime since no secret-key data is encrypted with it. The three beams: two at $\lambda \sim 850$ nm and one at $\lambda \sim 1550$ nm must be combined in a single beam to be transmitted by the same telescope. This is achieved by a 50/50 beamsplitter and a broadband pellicle beamsplitter. The three beams are then expanded and collimated to an approximate diameter of 40 mm.

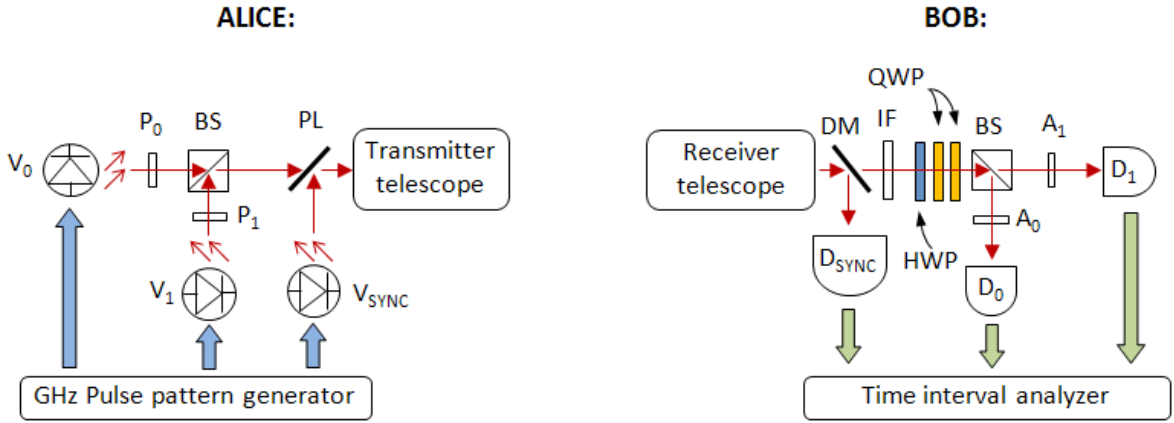

Fig. 1. Schematic of the QKD system. $V_0$, $V_1$ and $V_{SYNC}$ are laser diodes; $P_0$ and $P_1$ are high-extinction ratio polarizers; BS is a 50/50 beamsplitter; PL is a pellicle beamsplitter; DM is a dichroic mirror; IF is an interference filter; HWP and QWP are half and quarter waveplates respectively; $A_0$ and $A_1$ are polarization analyzers; $D_0$ and $D_1$ are single-photon detectors; $D_{sync}$ is an amplified photodetector.

The photons reaching the receiver are focused using a Schmidt-Cassegrain telescope and spectrally discriminated with a dichroic mirror, which transmits the 850 nm data beam, and reflects the 1550 nm synchronization beam. Photons from the 850 nm source are then split into two paths independently of their polarization. A polarizer is used in each channel to correctly select the photons of each state. The arrival times are then analyzed in a time interval analyzer.

## 3   Influence of sky background and turbulence

Alice and Bob analyze the error rate of a small subset of the photon sequence received by Bob in order to assess the security of a key transmission. The error rate, commonly referred to as quantum bit error rate (QBER), is defined as the number of incorrect bits over the total number of bits received by Bob. If the QBER is higher than a certain threshold, which depends on the protocol and implementation – 8 % for this system – the transmission is not considered as secure. Many factors influence the QBER, such as the noise of the single-photon detectors, polarization imperfections of the quantum states, but also atmospheric turbulence and solar background radiation, when operated during the day. As will be explained, these last two factors show effects that imply confronting mitigation solutions.

A combination of spatial, software and spectral filtering is usually implemented in order to reduce the background noise. In our case, a spectral filter with a bandwidth of less than 1 nm combined with spatial filtering by optical fibers with a core diameter of 62.5 µm was considered sufficient to reduce the background radiation to acceptable levels to enable fast key rate generation at times of the day when the radiation level is not too high (e.g. sunrise or sunset). However, when the sun radiation reached high intensities, around noon or at some time intervals when the sun was hitting directly the receiver, the error rate increases considerably to levels in the region of 5 % to 7 % causing a highly reduced secure key rate. Therefore, the receiver field of view (FOV) needs to be minimized as the main strategy to limit the background noise during the day.

Atmospheric turbulence is a random space-time distribution of the refractive index, due to air masses movements from thermal fluctuations, which affects the optical wavefront in different ways among which the most noticeable for QKD optical links are *beam spreading* and *beam wandering*. The former is caused by turbulent eddies that are small compared to the beam size and its main effect is an increase in the beam divergence. Beam wander, on the other hand, has its origin in turbulent eddies larger than the beam size resulting in random deflections of the laser beam. Both effects can be combined in the *long-term beam radius*, which models the effective size of the laser spot at the receiver as the result of divergence due to diffraction and beam spreading, and the displacement of the beam caused by beam wander over a long time period.

In any free-space lasercom system operating during daylight, a key strategy to limit the sky background reaching the detector is minimizing the FOV of the receiver's detector. This is especially important in high-speed QKD since the optical signal cannot be amplified and the background increases the error rate, due to the extremely sensitive detectors used at the receiver; consequently reducing the achievable bit rate. The spatial filtering to reject background is achieved by decreasing the diameter of the optical fiber connected to the detector, which directly reduces the FOV of the system. However, a narrow FOV is highly sensitive to the angle-of-arrival fluctuations originated by beam wander. These deflections cause the signal to focus in different places of the focal plane leading it to fall out of the optical fiber aperture with the consequent temporal interruptions in the transmission.

The longer the distance, the more noticeable the effect of the turbulence for a QKD system with no active tracking. In order to avoid this significant performance limitation, a tracking system needs to be implemented to compensate these fluctuations while allowing a more significant FOV reduction to limit the background noise.

# 4   High-speed tracking techniques: pre-compensation at the emitter and compensation at the receiver

There are various possible setups to achieve beam wander correction. All of them are based on variations of classical laser alignment systems, consisting in a fast-steering mirror (FSM), which is fed with data from a position-sensitive detector (PSD) closing the loop with a proportional-integral-derivative (PID) control. Beam wander is a fast phenomenon, with varying rates that can exceed a hundred Hz; therefore it is important to design the tracking system to accommodate for these fluctuation rates, especially the mechanical part (i.e. the fast-steering mirror). The optical paths carrying the quantum and timing information from Alice to Bob will be referred to as *data* and *sync channels* and that used to extract the information for beam compensation will be the *tracking channel*.

## *4.1  Pre-compensation at the emitter.*

Beam wander can be modeled as if it was originated from a tip-tilt variation of the laser beam at the transmitter or as an angle-of-arrival fluctuation at the receiver; these two approaches giving rise to two different mitigation techniques. If the long-term beam diameter at the receiver is larger than its aperture size, the compensation should be done at the transmitter. Otherwise, the link would suffer from big losses as the beam could fall out of the receiver aperture. Fig. 2 shows the suggested setup to implement this strategy. The goal is to pre-compensate in Alice the beam wander affecting the quantum channel, using a 'backwards' tracking channel consisting in a laser beam being transmitted from Bob to Alice. The measurements performed on the position of this 'backwards' tracking beam are analyzed in Alice and compensated for with fine movements of Alice's FSM. This method may result in a more complex setup involving an additional laser at a different wavelength from that of the quantum channel, in order to avoid possible back reflections that could couple into the single-photon detectors at the receiver, but it is valid for any relationship between the long-term beam diameter at the receiver's end and the receiver's aperture size. However, this strategy has a maximum distance of application, since for very long transmission paths, the changes in the atmosphere will be faster than the time involved in the pre-compensation (i.e., the temporal correlation between both atmospheric channels in opposite ways will be lost). This limit will depend on the parameters of the system, but since the time scale of beam wander is of the order of milliseconds, the distance limit would be in the hundreds of kilometers.

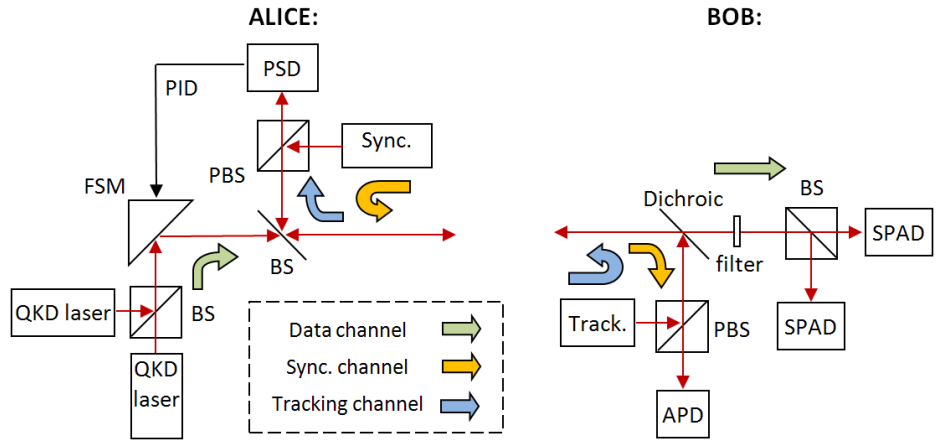

**Fig. 2. Pre-compensation at the emitter.** The data channel is the combined signal from two lasers and is reflected in Alice by a fast-steering mirror (FSM) which compensates for the fluctuations in the transmission channel by analyzing the received tracking channel from Bob with the control loop composed of the position sensitive detector (PSD) and the proportional integral derivative (PID). In Bob, the dichroic mirror reflects both the tracking and sync channels (although in opposite ways) and transmits the data channel to the single photon avalanche diodes (SPADs). The tracking and sync channels are polarized in orthogonal polarization and therefore are transmitted and reflected in opposite directions by a polarizing beamsplitter (PBS).

## 4.2 *Compensation at the receiver.*

In scenarios where the long-term beam diameter is not larger than the receiver aperture a simpler compensation is possible. This is typically the case for short-to-medium-range paths, (not beyond ~3 km in our system for a low-to-moderate turbulence regime, as will be discussed later). In this scenario the received spot always enters the telescope and no pre-compensation is thus needed in Alice. Compensation at the receiver (whose schematic can be seen in fig. 3) is therefore sufficient to correct for atmospheric fluctuations affecting the link. Furthermore, since this scheme is not bidirectional, unlike the previous one, the synchronization and the tracking channels can be combined in just one channel, therefore requiring only two lasers. This method is aimed at correcting the angle deviations of the received beam caused by turbulence effects by realigning the beam to a pre-established optimum position on a PSD where the system is perfectly aligned.

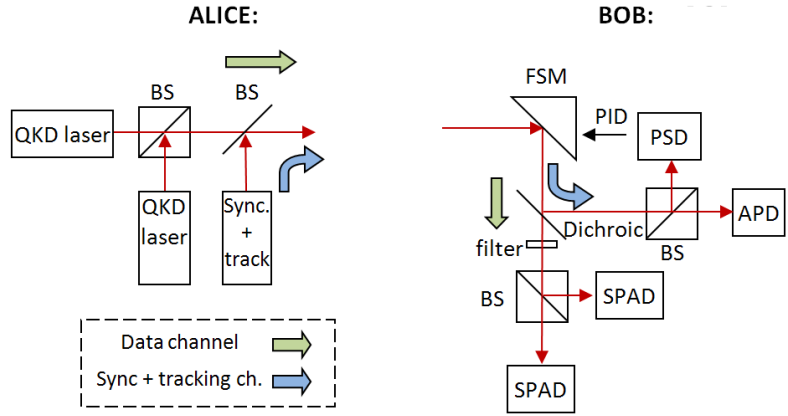

Fig. 3. Compensation at the receiver. Both the data and the sync+tracking channels are emitted from Alice and spectrally discriminated at the receiver by a dichroic mirror. The sync+tracking channel is deflected by a fast steering mirror (FSM) in Bob to a beamsplitter (BS) where a fraction of it is used for synchronization purposes (directed to an avalanche photodiode (APD)) and another fraction is used for tracking (directed to a position sensitive detector (PSD)).

Depending on the characteristics of the link, compensation at the receiver may be enough to mitigate beam wander. Otherwise, a pre-compensation at the emitter should be used instead. The type of compensation that should be used will depend on the optical design of Alice and Bob (mainly, the sizes of both Alice's output beam and Bob's collecting aperture), on the turbulence regime (characterized by the refractive-index structure constant $C_n^2$), and on the propagation distance. As a rule of thumb, the pre-compensation principle will always work, however, in order to simplify the system, the receiver compensation may be preferred. The main reason for this, as mentioned previously, is that the former setup involves transmitting a backwards tracking laser perfectly aligned with the quantum channel in such a way that both beams are transmitted through the exact same atmospheric path (although in opposite directions). This involves the use of an additional laser in the system and a very demanding alignment technique that could critically affect the system performance when not perfectly calibrated.

For a particular turbulence regime, the receiver-compensation setup could be extended to longer distances if a larger transmission aperture $D_T$ (i.e., a larger transmitted beam) and/or receiver aperture $D_R$ was used. This choice of apertures should be carefully designed, as an optimum value for both parameters exists. This optimum value corresponds to the minimum beam divergence caused by both diffraction and beam spreading and it will be different for each distance. Fig. 4 shows these optimum values for the apertures of transmitter and receiver at different turbulent regimes in a receiver-compensation configuration. For example, the receiver compensation configuration could be used to distances of up to 5 km, if the transmitter and receiver apertures were selected to be the optimum ones, that is 7.4 cm and 13.1 cm respectively (see fig. 4), assuming an average turbulent regime ($C_n^2 = 10^{-15}$ m$^{-2/3}$).

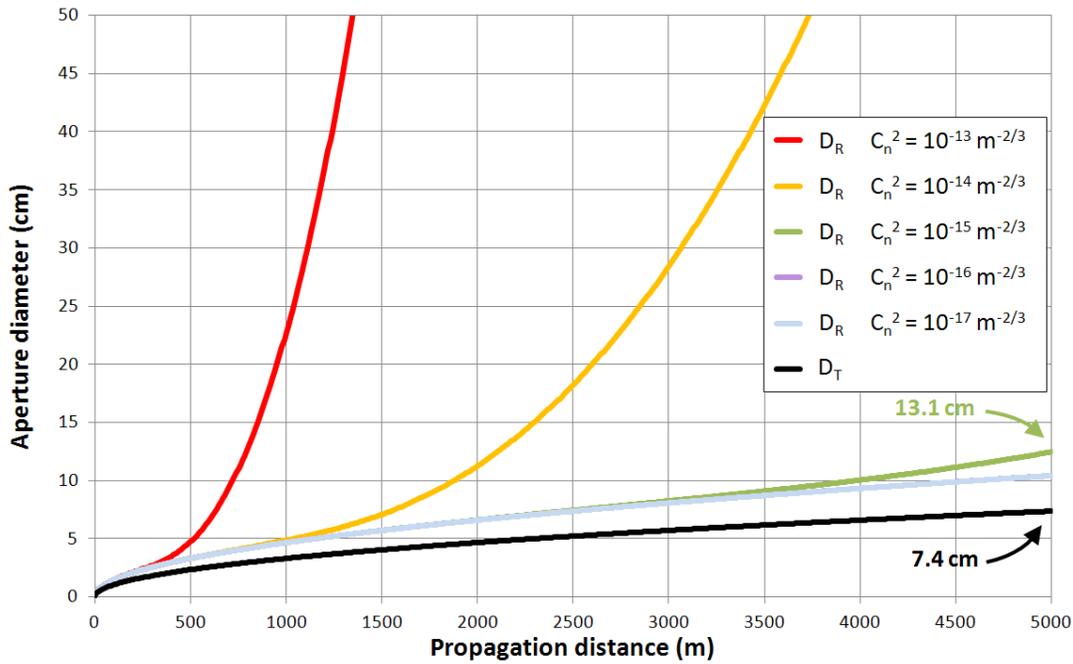

**Fig. 4. Optimum transmitter, $D_T$, and receiver, $D_R$, apertures versus propagation distance for different turbulent regimes in a receiver-compensation setup. $D_T$ corresponds to the maximum diameter that can be sent by the transmitter's telescope and $D_R$ the maximum beam collected by the receiver without incurring in losses.**

A useful way to choose the compensation strategy is to study the ratio of the receiver aperture to the received long-term diameter. If this ratio is greater than 1, then the beam always enters the receiver aperture and therefore the compensation can be made at the receiver end. Otherwise, the compensation should be made at the transmitter to avoid extra losses. For each turbulence regime, a maximum distance can be determined as a boundary between the two methods (see fig. 5). For this calculation, classic second-order turbulence fluctuation statistics [6] have been used, along with real data from the system described in section 2. It can be observed that at $C_n^2 = 10^{-15}$ m$^{-2/3}$ – which is often referred to as an average turbulence regime –, compensation at the receiver can be used for distances lower than 2.45 km. This is because in this region the long-term beam diameter at Bob is smaller than the receiver aperture, i.e., the ratio of receiver aperture to the long-term diameter is larger than 1 and we are at the top side of the graph represented in fig. 5. For longer distances than 2.45 km, pre-compensation at the emitter should be chosen, since the long-term beam diameter falls outside the receiver aperture. For a very strong turbulence regime, ($C_n^2 = 10^{-14}$ m$^{-2/3}$), compensation at Bob would be limited to a distance of less than 1.65 km, with pre-compensation being the more efficient mitigation strategy for longer distances. The case of $C_n^2 = 10^{-13}$ m$^{-2/3}$ is considered an extremely and unusual strong regime. Nevertheless, if such a turbulence was expected, 800 m would be the limit where compensation at the receiver should be used whereas the alternative method should be used for larger distances.

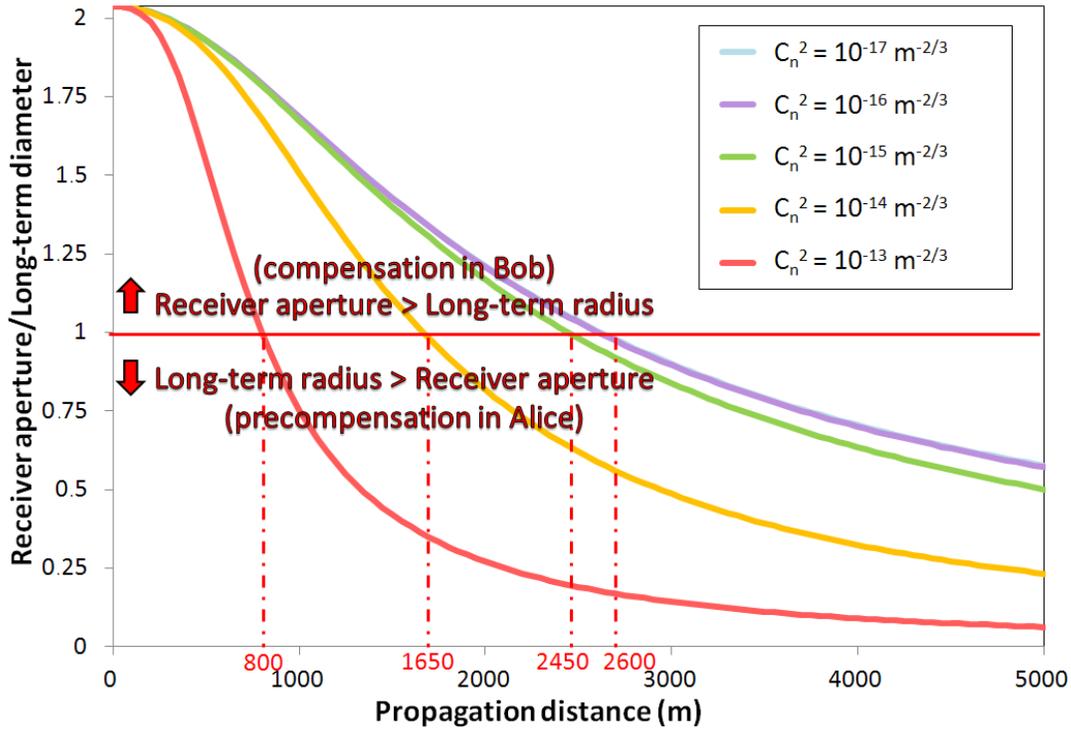

**Fig. 5.** Ratio of the receiver aperture to long-term beam diameter as a function of distance for several turbulence regimes at λ ~ 850 nm for a transmitted beam of 4 cm and a receiver aperture of 8 cm (system described in [4]).

# 5 Considerations for the experimental implementation of automatic tracking systems

In this section, we present some experimental considerations for the tracking system. Although these considerations will be valid for both proposed approaches, they are referred to the compensation-at-the-receiver setup due to the latter being chosen as the best tracking technique for the system in [4]. The reason for this was that this tracking technique is optimum for shorter distances. Coincidentally, the studied QKD system is also affected by a distance limitation due to the high penalization in bit rate with distance of the eavesdropping attacks considered in the calculation of the secret key rate (SKR) in [4].

A critical part of the tracking system is the optics that match the received beam size to the detector's active area for several reasons: first, the wandering range of the beam can be reduced if the beam is focused and its displacement is monitored in the focal plane. Moreover, in this position the FSM angular range is lower, which means the response can be faster, since a trade-off exists between the FSM maximum beam deflection and the fastest achievable response. The range of the FSM angular movements must therefore be minimized in order to reduce the time response of the tracking system. Our commercial off-the-shelf FSM has a time response of less than 10 milliseconds, which is sufficiently precise to correct most of the turbulence variations, which typically reach frequencies of up to 100 Hz.

Another reason for using well-focused beams in the correction is related to the quantum channel and aims to decrease the FOV and therefore reduce the solar background coupled into the QKD receiver. This reduction is achieved mainly by using a small-aperture single-mode optical fiber. The 62.5 μm optical fiber used in [4] could accommodate most of the turbulence-related beam displacements. However, a reduction of almost one order of magnitude in the optical fiber core diameter – from multi-mode to single-mode at λ ~ 1550 nm –, which could reduce considerably the background noise of the system, demands not only a good correction of the beam wander by the tracking technique, but also a very small spot size.

There is a critical consideration when implementing the proposed correction-at-the-receiver setup, which is the relative position of the detectors in both channels (sync+tracking and data channels) after the focusing optics, i.e., the distances from the point where the correction is made (FSM) to the position of each detector, SPAD or PSD. These distances are effectively $d_{SPAD}$ and $d_{PSD}$ in fig. 6 (left), since the distance from the FSM to the dichroic mirror is common for both channels. In fig. 6 (right), an optical simulation of the detected centroid displacement in each channel is shown versus the relative detector position after a beam has been corrected using the receiver correction setup shown in fig.6 (left). Since the correction is made using the PSD signal, only negligible displacements are observed in this channel (sync+tracking), whereas in the quantum data channel larger variations in the position of the centroid, which can take away the beam from the optical axis, were observed. However, making $d_{SPAD}$ equal to $d_{PSD}$, the relative variations in the centroid are reduced considerably. Therefore, it is essential that both detectors are placed exactly at the same relative position, or the data beam will not be efficiently corrected.

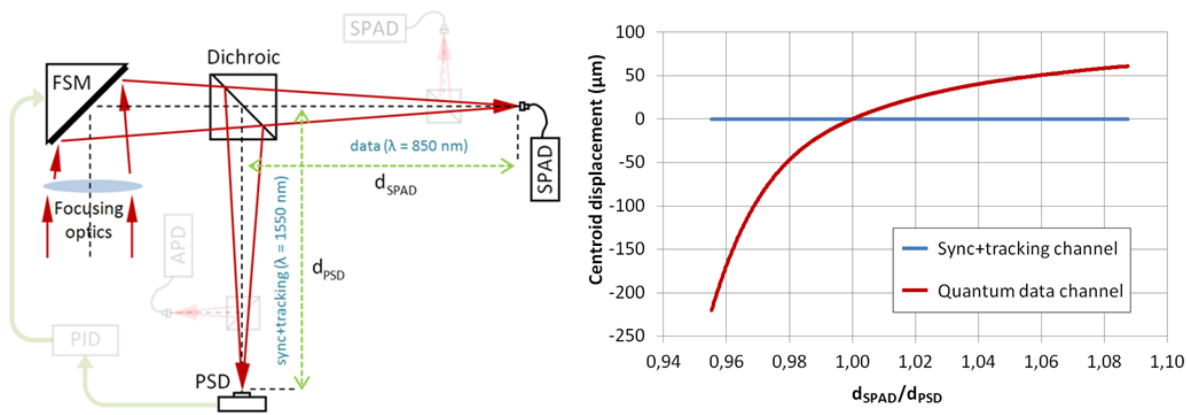

**Fig. 6. Beam centroid displacement (right) versus relative detector position in each channel – sync+tracking and data channels – when correcting a beam with the receiver correction setup (left).**

On the other hand, to achieve the smallest spot at the focal point in Bob, where the optical fiber is placed, it is important to use aberration-free optics. Ideally, aspheric optics should be used in order to minimize the focus size, but it is also important to reduce chromatic aberration since two different wavelengths have to reach the detectors placed exactly at the same relative distance. In practice, as both wavelengths are very far from each other, even achromatic optics will introduce a great amount of chromatic aberration, so either a single curved aspheric mirror or slightly different optics should be used in each channel to get the same effective focal length.

Another consideration for the experimental implementation of tracking systems in QKD is that due to the mentioned need of highly focused beams for the efficient coupling to optical fibers, the position sensitive detector used to monitor the spot displacements cannot be a quadrant sensor, as it is usually the case in most tracking systems. This kind of detector provides the best linear response, but they operate better with slightly defocused beams to avoid the gaps between adjacent cells. With state-of-the-art technology, this gap can be as small as tens of μm, but still adequate beams for these detectors are too large to be coupled into single-mode optical fiber at λ ~ 1550 nm. Thus, a lateral effect PSD was selected for the tracking system designed for the QKD system described here. This kind of sensors are much less common and more expensive than quadrant sensors, especially when fabricated in the InGaAs technology needed to detect the λ ~ 1550 nm tracking laser used for this system. Besides, they are not as linear as quadrant detectors, although using a sufficiently small area in the middle region of the sensor – which is the case when using highly focused beams, as it is the case described here –, generally avoids this problem, since only the linear region is used. On the other hand since they are built using one single active area without any gaps, the spot can be as small as desired, which is the goal in our system.

# 6 Preliminary experimental results and evaluation of the potential improvement in terms of QBER

In order to assess the validity of the proposed approach and potential improvement in key parameters of the QKD system, such as the QBER or SKR, some experimental measurements were performed. Fig. 7 shows the correction of beam wander in a 30-meter link between two buildings in Madrid under strong-turbulence regime. The experiment was carried out using a simplified version of the compensation-at-the-receiver configuration (fig. 3) with no telescopes being used in Alice nor Bob due to the short distance of the link. The beam diameter was also reduced (from 40 mm in the original system to 9 mm in the experiment) to increase the effect of beam wander, which is inversely proportional to the beam diameter.

Analyzing the data in fig. 7, we find that 90 % of the events fall within a distance from the optical axis of 0.55 μm at the focal plane when the correction was being applied, compared to 4.75 μm with no correction. Thus, the correction system improves this distance by a factor of ~9, which implies a focal area ~75 times smaller. The correction factor obtained from the experimental results was extended to a range of values between 5 and 10 to consider a wider and more realistic scenario of probabilities. This range of values was then used to estimate the potential improvement of the studied QKD system. In order to do this, we simulated the performance in terms of QBER and SKR versus distance using previous experimental data from our QKD system. In particular, data from background and signal photons taken from the 300-meter link described in [4] were adjusted to take into account the correction factor. For this simulation, the turbulence model in [6] was used and we assumed an atmospheric absorption of 2.048 dB/km, calculated with MODTRAN (urban aerosol and visibility of 5 km). We also modelled the corrected long-term diameter at the focus of the receiver to be coupled to an optical fiber with the same core size in order to limit the background noise. The results can be seen in fig. 8, where the improvements in QBER and SKR are shown under a strong turbulence regime (with a maximum applicable distance of 800 m, as established in fig. 5). The experimental data were taken under strong turbulence, so this was the regime selected in the simulation to show the best correction of the system due to the stronger effect of beam wander.

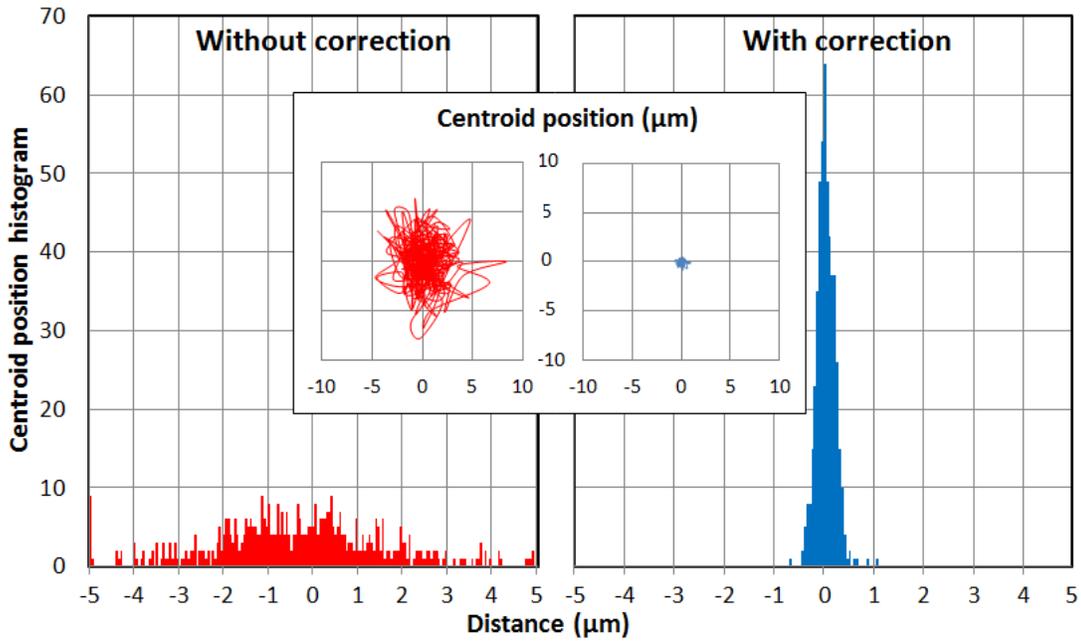

Fig. 7. Histogram of the centroid position of the beam at the receiver's focal plane recorded over a period of 15 seconds, with and without beam wander correction. Embedded, a graph showing the actual centroid position at the focal plane for the same time interval.

Since the correction factor was simulated for a range of values, the corrected data appears to be a thick line in fig. 8. The QBER is significantly reduced (up to 80 %, especially for stronger turbulent regimes), implying a considerable increase in the SKR (up to three orders of magnitude under strong turbulence). The QBER and SKR were also calculated for the medium and weak turbulent regimes. The QBER decreased 50 % and 10 % for the two lower regimes and the SKR increased up to two orders of magnitude for the medium regime and one for the weak regime. It should also be stressed that beam wander correction not only allows achieving higher key rates due to lower values of QBER, but also enables a secure transmission over longer distances since it lowers the maximum QBER limit of 8 % of the studied QKD system.

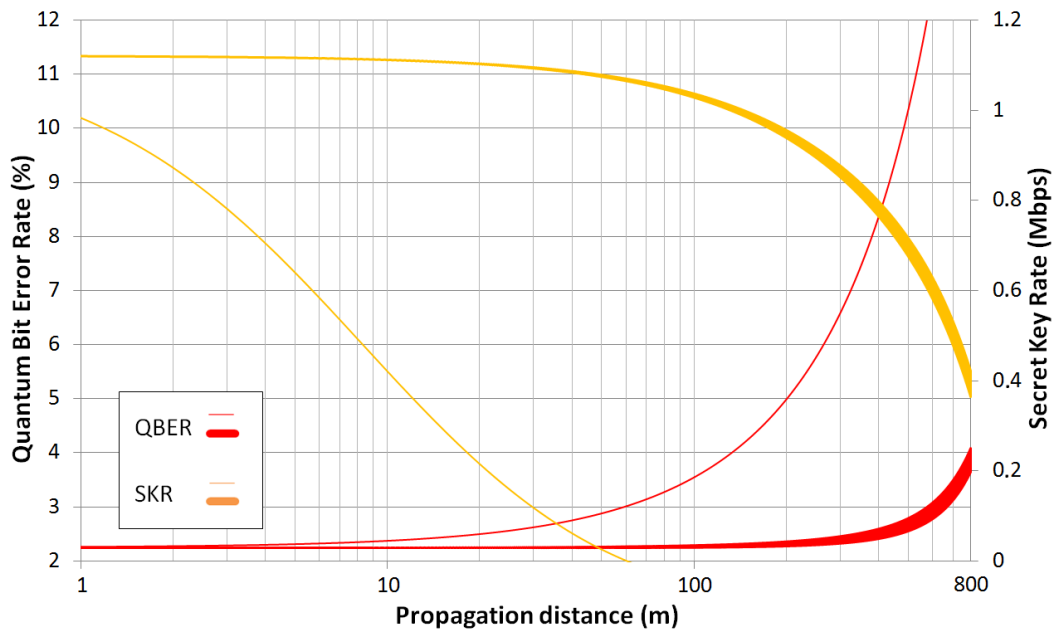

Fig. 8. Quantum Bit Error Rate and Secret Key Rate as a function of the propagation distance with (thick line) and without (thin line) beam wander correction under a strong turbulent regime ($C_n^2 = 10^{-13}$ m$^{-2/3}$).

# 7 Conclusions

In this paper, the implementation of a high-speed tracking system for a free-space QKD system correcting for atmospheric turbulences in urban environment has been presented and discussed. Controlling critical parameters such as solar background and turbulence is essential to achieve high secure key exchange rates in a QKD link. Reducing the FOV decreases the background noise coupled in a QKD system, which is necessary to keep the QBER down. However, to do so while avoiding losses in the system due to the position of the beam moving away from the optimum position of alignment, an efficient and reliable turbulence compensation technique must be used. The compensation of turbulence effects through two different approaches has been proposed, each technique being more efficient depending on the turbulent regime and the propagation distance. The compensation technique chosen for our system - compensation at the receiver - was discussed along with some technical parameters such as the transmitter and receivers' optimum apertures that should be taken into consideration for the implementation of the tracking system. Some considerations regarding the experimental implementation of the tracking system, necessary for a more efficient compensation, have been addressed, such as the use of focused beams, an equal relative position of the detectors in the tracking setup, the use of aspheric and achromatic optics, along with lateral effect PSD instead of quadrant detectors to implement the compensation. Finally, preliminary experimental results from a simplified correction system have been presented and applied to estimate the potential improvement in the performance of the QKD system in terms of QBER and SKR. This improvement involves a significant decrease of the QBER providing higher secure key rates in general and allowing key transmission at previously forbidden distances.

## *Acknowledgments*

We thank the project TEC2012-35673 from the Ministerio de Economía y Competitividad.

6. L. C. Andrews and R. L. Phillips, *Laser Beam Propagation Through Random Media*, SPIE Press Monograph Vol. PM152 (2005).

*Authors*

**Alberto Carrasco-Casado** received his B.S. degree in telecommunication engineering from Málaga University in 2005. He received both the M.S. degree in telecommunication engineering and M.S. degree in space research from Alcalá University in 2008 and 2009, respectively. He is currently pursuing his Ph.D. degree at Spanish National Research Council in Madrid, Spain. His research interests focuses on free-space optical communications.

**Veronica Fernandez** received the B.Sc. degree (hons.) in physics with electronics from the University of Seville, Seville, Spain, in 2002 and the Ph.D. degree in physics from Heriot-Watt University, Edinburgh, U.K., in 2006. In 2007, she joined the Cryptography and Information Security group at the Spanish National Research Council (CSIC), where she was granted a tenured scientist position in 2009. She has coauthored several journal articles and conference publications on fiber and free-space based quantum key distribution.

**Natalia Denisenko** received her B.S. degree in electrical and electronic engineering from Moscow Power Engineering Institute in 1975. She worked at Laser Physics Laboratory (Paris University, France) between 1975 and 1977, at Materials Physics Institute (Spanish National Research Council, Madrid, Spain) between 1979 and 1982 and at Ultrasonic and Acoustic Technologies Laboratory (National Research Council, Rome, Italy) between 1982 and 1985. In 1986, she joined the Spanish National Research Council (Madrid, Spain).